\documentclass[a4paper]{JACoW-GSI}
\usepackage{graphicx}
\usepackage{url}

\def\kbar{\overline{K}{}^{\,0}}
\def\dbar{\overline{D}{}^{\,0}}

\def\cp{$CP$}
\def\cpv{$CPV$}
\def\ra{\!\rightarrow\!}

\def\dklnu{$D^0\ra K^+\ell^-\nu$}
\def\dkpi{$D^0\ra K^+\pi^-$}

\def\dkkpp{$D^0\ra K^+K^-/\pi^+\pi^-$}
\def\dkspp{$D^0\ra K^0_S\,\pi^+\pi^-$}
\def\dkkks{$D^0\ra K^+K^-K^0_S$}
\def\ycp{$y^{}_{\rm CP}$}

\def\meve{~MeV}

\def\babar{Babar}

\def\simge{\mathrel{%
   \rlap{\raise 0.511ex \hbox{$>$}}{\lower 0.511ex \hbox{$\sim$}}}}
\def\simle{\mathrel{
   \rlap{\raise 0.511ex \hbox{$<$}}{\lower 0.511ex \hbox{$\sim$}}}}

\begin{document}

\title{
\vspace*{-0.3in}
\begin{flushright}
{\large {\rm UCHEP-09-03}}
\end{flushright}
{\boldmath $D^0$-$\dbar$ Mixing and \cp\ Violation:
HFAG Combination of Parameters}
}

\author[1]{
\vspace*{-0.15in}
A. J. Schwartz\thanks{alan.j.schwartz@uc.edu} \\
(representing the HFAG charm group)}
\affil{Physics Department, University of Cincinnati, Cincinnati, OH 45221, USA
}

\maketitle

\abstract{We present the most recent set of world averages for 
$D^0$-$\dbar$ mixing and \cp\ violation parameters, as obtained 
by the Heavy Flavor Averaging Group from a global fit to various 
measurements. The values obtained for the mixing parameters when 
allowing for \cp\ violation are 
$x= (0.98\,^{+0.24}_{-0.26})\%$ and 
$y= (0.83\,\pm 0.16)\%$; the significance of mixing is $10.2\sigma$. 
There is no evidence for \cp\ violation at the current level 
of sensitivity.}

\section{Introduction}

In 2006, the Heavy Flavor Averaging Group (HFAG)~\cite{hfag-main}
convened a new subgroup to calculate world average (WA) values of 
charm mixing and \cp\ violation (\cpv) parameters~\cite{hfag-charm}. 
Since that time, $D^0$-$\dbar$ mixing has been observed, and a wealth 
of mixing and \cpv\ results have appeared. The HFAG charm group has 
calculated several sets of WA values, updating old averages as new 
results have become available. This paper presents the most recent 
set of averages, i.e., those based on results that appeared in 
preprint form by the summer of~2009.

Mixing in the $B^0$ and $B^0_s$ heavy flavor systems is 
governed by the short-distance box diagram. In the $D^0$ system,
however, this diagram is doubly-Cabibbo-suppressed (relative to 
amplitudes dominating the decay width) and also GIM-suppressed.
Thus, the short-distance mixing rate is tiny, and $D^0$-$\dbar$ 
mixing is expected to be dominated by long-distance 
processes. These are difficult to calculate reliably, and 
theoretical estimates for $D^0$-$\dbar$ mixing range over 
2-3 orders of magnitude~\cite{BigiUraltsev,Petrov}.

The decay rates for $D^0\ra f$ and $\dbar\ra\bar{f}$ are, respectively,
\begin{eqnarray}
\frac{dN^{}_{D^0}}{dt} & \!\!\propto\!\! & 
e^{-\overline{\Gamma}\,t}\ \left\{ 
R^+ + \left|\frac{q}{p}\right|\sqrt{R^+}\,\times \right. \nonumber \\
 & & \hskip0.4in 
(y'\cos\phi - x'\sin\phi )(\overline{\Gamma}t) +  \nonumber \\
 & & \hskip0.7in \left. \left|\frac{q}{p}\right|^2 
\frac{(x'^2 + y'^2)}{4}(\overline{\Gamma}\,t)^2 \right\} \\ \nonumber \\
\frac{dN^{}_{\dbar}}{dt} & \!\!\propto\!\! & 
e^{-\overline{\Gamma}\,t}\ \left\{ R^- + 
\left|\frac{p}{q}\right|\sqrt{R^-}\,\times \right. \nonumber \\ 
 & & \hskip0.4in 
(y'\cos\phi + x'\sin\phi )(\overline{\Gamma}t) +  \nonumber \\
 & & \hskip0.7in \left. \left|\frac{p}{q}\right|^2 
\frac{(x'^2 + y'^2)}{4}(\overline{\Gamma}\,t)^2 \right\}\!.
\end{eqnarray}
In these expressions,
$x' = x\cos\delta + y\sin\delta$ and 
$y' = y\cos\delta - x\sin\delta$, where
$x= (M^{}_2-M^{}_1)/\overline{\Gamma}$ and 
$y= (\Gamma^{}_2-\Gamma^{}_1)/(2\overline{\Gamma})$
are mixing parameters, and $\delta$ is the strong phase 
difference between amplitudes 
${\cal A}(\dbar\ra f)$ and ${\cal A}(D^0\ra f)$.
Parameters $M^{}_1$, $M^{}_2$, $\Gamma^{}_1$, and $\Gamma^{}_2$
are the masses and decay widths of the mass eigenstates
$|D^{}_1\rangle\equiv p|D^0\rangle + q|\dbar\rangle$ and
$|D^{}_2\rangle\equiv p|D^0\rangle - q|\dbar\rangle$, and
$\overline{\Gamma} = (\Gamma^{}_1 + \Gamma^{}_2)/2$.
Our convention is $CP|D^0\rangle = -|\dbar\rangle$ such that
for $q\!=\!p$, $D^{}_1$ is $CP$-odd and $D^{}_2$ is $CP$-even.
The parameters 
$R^+ = |{\cal A}(D^0\ra f)/{\cal A}(\dbar\ra f)|^2$,
$R^- = |{\cal A}(\dbar\ra\bar{f})/{\cal A}(D^0\ra\bar{f})|^2$,
and $\phi = {\rm Arg}(q/p)$.

To obtain WA values of $x,y,\delta,|q/p|$, and $\phi$, we perform 
a global fit to 28 measured observables. These observables are from 
measurements of \dklnu, \dkkpp, \dkpi, $D^0\ra K^+\pi^-\pi^0$, 
\dkspp, and \dkkks\ decays~\cite{charge-conjugates}, and from
double-tagged branching fractions measured in
$e^+e^-\ra \psi(3770)\ra DD$ reactions.
To fit these observables, we must include 
an additional strong phase $\delta^{}_{K\pi\pi}$ (see below).
For $D^0\ra K^+\pi^-$ decays, we combine $R^+$ 
and $R^-$ into parameters 
$R^{}_D\equiv (R^+ + R^-)/2$ and 
$A^{}_D\equiv (R^+ - R^-)/(R^+ + R^-)$.
Correlations among observables are accounted for by using 
covariance matrices provided by the experimental collaborations. 

With the exception of the $\psi(3770)\ra DD$ measurements, all 
methods identify the flavor of the $D^0$ or $\dbar$ when produced by 
reconstructing the decay $D^{*+}\ra D^0\pi^+$ or $D^{*-}\ra\dbar\pi^-$; 
the charge of the accompanying pion identifies the $D$ flavor. For signal 
decays, $M^{}_{D^*}-M^{}_{D^0}-M^{}_{\pi^+}\equiv Q\approx 6$\meve, 
which is relatively close to the threshold. Thus, analyses typically
require that the reconstructed $Q$ be small to suppress backgrounds. 
For time-dependent measurements, the $D^0$ decay time is 
calculated as $(\ell/p)\times M^{}_{D^0}$, where $\ell$ is
the distance between the $D^*$ and $D^0$ decay vertices and 
$p$ is the $D^0$ momentum. The $D^*$ vertex position is 
taken to be either the primary vertex position
($\bar{p}p$ experiments) or else is calculated 
from the intersection of the $D^0$ momentum 
vector with the beam-spot profile ($e^+e^-$ experiments).

\section{\boldmath Input Observables}

The global fit determines central values and errors for
$x,\,y,\,\delta,\,R^{}_D, A^{}_D,\,|q/p|,\,\phi$, and 
$\delta^{}_{K\pi\pi}$ using a $\chi^2$ statistic. 
Parameters $x$ and $y$ govern mixing, and
parameters $A^{}_D$, $|q/p|$, and $\phi$ govern~\cpv.
The parameter $\delta^{}_{K\pi\pi}$ is the strong phase 
difference between amplitudes ${\cal A}(\dbar\ra K^+\pi^-\pi^0)$ 
and ${\cal A}(D^0\ra K^+\pi^-\pi^0)$ evaluated at 
$M^{}_{K^+\pi^-}=M^{}_{K^*(890)}$.

\begin{figure}
\includegraphics[width=84mm]{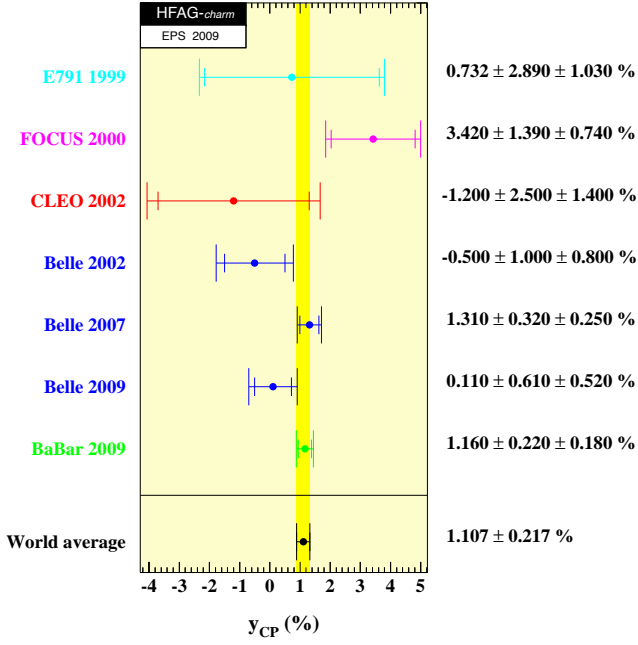}
\caption{WA value of $y^{}_{CP}$ as calculated from
$D^0\ra K^+K^-$, $D^0\ra\pi^+\pi^-$, and $D^0\ra K^+K^-K^0_S$ 
measurements~\cite{ycp_references}.}
\label{fig:ycp}
\end{figure}

All input values are listed in Table~\ref{tab:observables}. 
The values for observables $R^{}_M=(x^2+y^2)/2$~\cite{semi_references},
$y^{}_{CP}$~\cite{ycp_references}, and $A^{}_\Gamma$~\cite{ycp_references}
are HFAG WA values~\cite{hfag_charm}. They are calculated
as weighted averages of measurements, taking into account 
correlations among systematic errors and sometimes also 
statistical errors. As an example, the weighted average 
for $y^{}_{CP}$ is shown in Fig.~\ref{fig:ycp}.
The values of observables from \dkspp\ decays~\cite{kspp_references}
for no-\cpv\ are HFAG WA values~\cite{hfag_charm}, but
for the \cpv-allowed case only Belle measurements are available.
The \dkpi\ results used~\cite{kpi_references} are from Belle,
\babar, and CDF, as these results have much greater precision 
than earlier ones.
The $D^0\ra K^+\pi^-\pi^0$ results are from \babar~\cite{knpi_references}, 
and the $\psi(3770)\ra DD$ results are from CLEOc~\cite{cleoc}.

\begin{table*}
\renewcommand{\arraystretch}{1.3}
\begin{center}
\caption{\label{tab:observables}
Input values used for the global fit, from
Refs.~\cite{semi_references,ycp_references,
kspp_references,kpi_references,knpi_references,cleoc}.
}
\footnotesize
\begin{tabular}{ccl}
\hline\hline
\textbf{Observable} & \textbf{Value} & \textbf{Comment} \\
\hline
\begin{tabular}{c}
 $y^{}_{CP}$  \\
 $A^{}_{\Gamma}$
\end{tabular} & 
$\begin{array}{c}
(1.107\pm 0.217)\% \\
(0.123\pm 0.248)\% 
\end{array}$   &
\begin{tabular}{l}  
WA $D^0\ra K^+K^-/\pi^+\pi^-$ and $D^0\ra K^+K^-K^0_S$ results~\cite{hfag_charm}
\end{tabular} \\
\hline
\begin{tabular}{c}
$x$ (no \cpv) \\
$y$ (no \cpv) \\
$|q/p|$ (no direct \cpv) \\
$\phi$ (no direct \cpv) 
\end{tabular} & 
\begin{tabular}{c}
 $(0.811\pm 0.334)\%$ \\
 $(0.309\pm 0.281)\%$ \\
 $0.95\pm 0.22^{+0.10}_{-0.09}$ \\
 $(-0.035\pm 0.19\pm 0.09)$ rad
\end{tabular} &
\begin{tabular}{l}
No \cpv: \\
WA $D^0\ra K^0_S\,\pi^+\pi^-$ results~\cite{hfag_charm}
\end{tabular} \\ 
 & & \\
\begin{tabular}{c}
$x$ \\
$y$ \\
$|q/p|$ \\
$\phi$  
\end{tabular} & 
\begin{tabular}{c}
 $(0.81\pm 0.30^{+0.13}_{-0.17})\%$ \\
 $(0.37\pm 0.25^{+0.10}_{-0.15})\%$ \\
 $0.86\pm 0.30^{+0.10}_{-0.09}$ \\
 $(-0.244\pm 0.31\pm 0.09)$ rad
\end{tabular} &
\begin{tabular}{l}
\cpv-allowed: \\
Belle $D^0\ra K^0_S\,\pi^+\pi^-$ results. Correlation coefficients: \\
\hskip0.30in $\left\{ \begin{array}{cccc}
 1 &  -0.007 & -0.255\alpha & 0.216  \\
 -0.007 &  1 & -0.019\alpha & -0.280 \\
 -0.255\alpha &  -0.019\alpha & 1 & -0.128\alpha  \\
  0.216 &  -0.280 & -0.128\alpha & 1 
\end{array} \right\}$ \\
Note: $\alpha=(|q/p|+1)^2/2$ is a variable transformation factor
\end{tabular} \\
\hline
 $R^{}_M$  & $(0.0130\pm 0.0269)\%$  &  
WA $D^0\ra K^+\ell^-\nu$ results~\cite{hfag_charm} \\
\hline
\begin{tabular}{c}
$x''$ \\ 
$y''$ 
\end{tabular} &
\begin{tabular}{c}
$(2.61\,^{+0.57}_{-0.68}\,\pm 0.39)\%$ \\ 
$(-0.06\,^{+0.55}_{-0.64}\,\pm 0.34)\%$ 
\end{tabular} &
\begin{tabular}{l}
\babar\ $D^0\ra K^+\pi^-\pi^0$ result. Correlation coefficient $=-0.75$. \\
Note: $x'' \equiv x\cos\delta^{}_{K\pi\pi} + y\sin\delta^{}_{K\pi\pi}$, 
$y'' \equiv y\cos\delta^{}_{K\pi\pi} - x\sin\delta^{}_{K\pi\pi}$.
\end{tabular} \\
\hline
\begin{tabular}{c}
$R^{}_M$ \\
$y$ \\
$R^{}_D$ \\
$\sqrt{R^{}_D}\cos\delta$  
\end{tabular} & 
\begin{tabular}{c}
 $(0.199\pm 0.173\pm 0.0)\%$ \\
 $(-5.207\pm 5.571\pm 2.737)\%$ \\
 $(-2.395\pm 1.739\pm 0.938)\%$ \\
 $(8.878\pm 3.369\pm 1.579)\%$ 
\end{tabular} &
\begin{tabular}{l}
CLEOc results from ``double-tagged'' branching fractions \\
measured in $\psi(3770)\ra DD$ decays. Correlation coefficients: \\
\hskip0.30in $\left\{ \begin{array}{cccc}
 1 &  -0.0644 &  0.0072 &  0.0607 \\
 -0.0644 & 1 & -0.3172 & -0.8331 \\
 0.0072 & -0.3172 & 1 & 0.3893 \\
 0.0607 & -0.8331 & 0.3893 & 1 
\end{array} \right\}$ \\
Note: the only external input to these fit results are \\
branching fractions.
\end{tabular} \\
\hline
\begin{tabular}{c}
$R^{}_D$ \\
$x'^{2+}$ \\
$y'^+$ 
\end{tabular} & 
\begin{tabular}{c}
 $(0.303\pm 0.0189)\%$ \\
 $(-0.024\pm 0.052)\%$ \\
 $(0.98\pm 0.78)\%$ 
\end{tabular} &
\begin{tabular}{l}
\babar\ $D^0\ra K^+\pi^-$ results. Correlation coefficients: \\
\hskip0.30in $\left\{ \begin{array}{ccc}
 1 &  0.77 &  -0.87 \\
0.77 & 1 & -0.94 \\
-0.87 & -0.94 & 1 
\end{array} \right\}$
\end{tabular} \\
\hline
\begin{tabular}{c}
$A^{}_D$ \\
$x'^{2-}$ \\
$y'^-$ 
\end{tabular} & 
\begin{tabular}{c}
 $(-2.1\pm 5.4)\%$ \\
 $(-0.020\pm 0.050)\%$ \\
 $(0.96\pm 0.75)\%$ 
\end{tabular} &
\begin{tabular}{l}
\babar\ $D^0\ra K^+\pi^-$ results; correlation coefficients same as above.
\end{tabular} \\
\hline
\begin{tabular}{c}
$R^{}_D$ \\
$x'^{2+}$ \\
$y'^+$ 
\end{tabular} & 
\begin{tabular}{c}
 $(0.364\pm 0.018)\%$ \\
 $(0.032\pm 0.037)\%$ \\
 $(-0.12\pm 0.58)\%$ 
\end{tabular} &
\begin{tabular}{l}
Belle $D^0\ra K^+\pi^-$ results. Correlation coefficients: \\
\hskip0.30in $\left\{ \begin{array}{ccc}
 1 &  0.655 &  -0.834 \\
0.655 & 1 & -0.909 \\
-0.834 & -0.909 & 1 
\end{array} \right\}$
\end{tabular} \\
\hline
\begin{tabular}{c}
$A^{}_D$ \\
$x'^{2-}$ \\
$y'^-$ 
\end{tabular} & 
\begin{tabular}{c}
 $(2.3\pm 4.7)\%$ \\
 $(0.006\pm 0.034)\%$ \\
 $(0.20\pm 0.54)\%$ 
\end{tabular} &
\begin{tabular}{l}
Belle $D^0\ra K^+\pi^-$ results; correlation coefficients same as above.
\end{tabular} \\
\hline
\begin{tabular}{c}
$R^{}_D$ \\
$x'^{2}$ \\
$y'$ 
\end{tabular} & 
\begin{tabular}{c}
 $(0.304\pm 0.055)\%$ \\
 $(-0.012\pm 0.035)\%$ \\
 $(0.85\pm 0.76)\%$ 
\end{tabular} &
\begin{tabular}{l}
CDF $D^0\ra K^+\pi^-$ results. Correlation coefficients: \\
\hskip0.30in $\left\{ \begin{array}{ccc}
 1 &  0.923 &  -0.971 \\
0.923 & 1 & -0.984 \\
-0.971 & -0.984 & 1 
\end{array} \right\}$
\end{tabular} \\
\hline\hline
\end{tabular}
\end{center}
\end{table*}

The relationships between the observables and the fitted
parameters are listed in Table~\ref{tab:relationships}. 
For each set of correlated observables, we construct a
difference vector~$\vec{V}$. For example, 
$\vec{V}=(\Delta x,\Delta y,\Delta |q/p|,\Delta \phi)$
for $D^0\ra K^0_S\,\pi^+\pi^-$ decays, where $\Delta$ 
represents the difference between the measured value 
and the fitted parameter value. The contribution of 
a set of observables to the $\chi^2$ is calculated as 
$\vec{V}\cdot (M^{-1})\cdot\vec{V}^T$, where $M^{-1}$ 
is the inverse of the covariance matrix for the 
measurement. All covariance matrices used 
are listed in Table~\ref{tab:observables}.

\begin{table*}
\renewcommand{\arraystretch}{1.3}
\begin{center}
\caption{\label{tab:relationships}
Left: decay modes used to determine the parameters 
$x,\,y,\,\delta,\,R^{}_D,\,A^{}_D,\,|q/p|,\,\phi$, and
$\delta^{}_{K\pi\pi}$.
Middle: observables measured for each decay mode. Right: the 
relationships between the observables and the fitted parameters.}
\footnotesize
\begin{tabular}{l|c|l}
\hline\hline
\textbf{Decay Mode} & \textbf{Observables} & \textbf{Relationship} \\
\hline
$D^0\ra K^+K^-/\pi^+\pi^-/K^+K^-K^0_S$  & 
\begin{tabular}{c}
 $y^{}_{CP}$  \\
 $A^{}_{\Gamma}$
\end{tabular} & 
$\begin{array}{c}
2y^{}_{CP} = 
\left(\left|q/p\right|+\left|p/q\right|\right)y\cos\phi\ -\ 
\left(\left|q/p\right|-\left|p/q\right|\right)x\sin\phi \\
2A^{}_\Gamma = 
\left(\left|q/p\right|-\left|p/q\right|\right)y\cos\phi\ -\ 
\left(\left|q/p\right|+\left|p/q\right|\right)x\sin\phi
\end{array}$   \\
\hline
$D^0\ra K^0_S\,\pi^+\pi^-$ & 
$\begin{array}{c}
x \\ 
y \\ 
|q/p| \\ 
\phi
\end{array}$ &   \\ 
\hline
$D^0\ra K^+\ell^-\nu$ & $R^{}_M$  & $R^{}_M = (x^2 + y^2)/2$ \\
\hline
\hskip-0.10in
\begin{tabular}{l}
$D^0\ra K^+\pi^-\pi^0$ \\
(Dalitz plot analysis)
\end{tabular} & 
$\begin{array}{c}
x'' \\ 
y''
\end{array}$ &
$\begin{array}{l}
x'' = x\cos\delta^{}_{K\pi\pi} + y\sin\delta^{}_{K\pi\pi} \\ 
y'' = y\cos\delta^{}_{K\pi\pi} - x\sin\delta^{}_{K\pi\pi}
\end{array}$ \\
\hline
\hskip-0.10in
\begin{tabular}{l}
``Double-tagged'' branching fractions \\
measured in $\psi(3770)\ra DD$ decays
\end{tabular} & 
$\begin{array}{c}
R^{}_M \\
y \\
R^{}_D \\
\sqrt{R^{}_D}\cos\delta
\end{array}$ &   $R^{}_M = (x^2 + y^2)/2$ \\
\hline
$D^0\ra K^+\pi^-$ &
$\begin{array}{c}
R^+,\ R^- \\
x'^{2+},\ x'^{2-} \\
y'^+,\ y'^-
\end{array}$ & 
$\begin{array}{l}
R^{}_D = (R^+ + R^-)/2 \\
A^{}_D = (R^+ - R^-)/(R^+ + R^-)  \\ \\
x' = x\cos\delta + y\sin\delta \\ 
y' = y\cos\delta - x\sin\delta \\
A^{}_M\equiv (|q/p|^4-1)/(|q/p|^4+1) \\
x'^\pm = [(1\pm A^{}_M)/(1\mp A^{}_M)]^{1/4}(x'\cos\phi\pm y'\sin\phi) \\
y'^\pm = [(1\pm A^{}_M)/(1\mp A^{}_M)]^{1/4}(y'\cos\phi\mp x'\sin\phi) \\
\end{array}$ \\
\hline\hline
\end{tabular}
\end{center}
\end{table*}

\section{\boldmath Fit results}

The global fit uses MINUIT with the MIGRAD minimizer, and
all errors are obtained from MINOS. Three separate fits are 
performed: {\it (a)}\ assuming \cp\ conservation ($A^{}_D$ 
and $\phi$ are fixed to zero, $|q/p|$ is fixed to one);
{\it (b)}\ assuming no direct \cpv\ ($A^{}_D$ is 
fixed to zero); and
{\it (c)}\ allowing full \cpv\ (all parameters floated). 
Results from the first and last fits are listed 
in Table~\ref{tab:results}. For the \cpv-allowed fit,
individual contributions to the $\chi^2$ are listed 
in Table~\ref{tab:results_chi2}. The total $\chi^2$ 
is 26.3 for $28-8=20$ degrees of freedom; this 
corresponds to a confidence level of~0.16.

\begin{table}
\renewcommand{\arraystretch}{1.4}
\begin{center}
\caption{\label{tab:results}
Results of the global fit for the cases of no \cpv\ and
all-\cpv-allowed.}
\footnotesize
\begin{tabular}{p{1.2cm}| p{1.7cm} p{1.9cm} p{1.9cm}}
\hline\hline
\textbf{Parameter} & \multicolumn{1}{c}{\textbf{\boldmath No \cpv}} 
& \multicolumn{1}{c}{\textbf{\boldmath \cpv-allowed}} & 
\multicolumn{1}{c}{\textbf{\boldmath \cpv\,95\% CL}}  \\
\hline
$\begin{array}{c}
x\ (\%) \\ 
y\ (\%) \\ 
\delta\ (^\circ) \\ 
\delta^{}_{K\pi\pi}\ (^\circ)  \\
R^{}_D\ (\%) \\ 
A^{}_D\ (\%) \\ 
|q/p| \\ 
\phi\ (^\circ) 
\end{array}$ & 
$\begin{array}{c}
0.99\,^{+0.24}_{-0.25} \\
0.81\,\pm 0.16 \\
25.2\,^{+9.6}_{-9.9} \\
13.5\,^{+20.2}_{-22.1} \\
0.336\,\pm 0.008 \\
- \\
- \\
- 
\end{array}$ &
$\begin{array}{c}
0.98\,^{+0.24}_{-0.26} \\
0.83\,\pm 0.16 \\
26.4\,^{+9.6}_{-9.9} \\
14.8\,^{+20.2}_{-22.1} \\
0.337\,\pm 0.009 \\
-2.2\,\pm 2.4 \\
0.87\,^{+0.17}_{-0.15} \\
-8.5\,^{+7.4}_{-7.0} 
\end{array}$ &
$\begin{array}{c}
 \mbox{[0.46, 1.44]} \\
 \mbox{[0.51, 1.14]} \\
 \mbox{[5.9, 45.8]} \\
 \mbox{[-30.3, 53.8]} \\
 \mbox{[0.320, 0.353]} \\
 \mbox{[-6.9, 2.6]} \\
 \mbox{[0.60, 1.22]} \\
 \mbox{[-22.1, 6.3]} 
\end{array}$ \\
\hline\hline
\end{tabular}
\end{center}
\end{table}

\begin{table}
\renewcommand{\arraystretch}{1.4}
\begin{center}
\caption{\label{tab:results_chi2}
Contributions to the $\chi^2$ (\cpv-allowed fit).}
\footnotesize
\begin{tabular}{l|rr}
\hline\hline
\textbf{Observable} & \textbf{\boldmath $\chi^2$} & \textbf{\boldmath $\sum\chi^2$} \\
\hline
$y^{}_{CP}$                      & 1.85 & 1.85 \\
$A^{}_\Gamma$                    & 0.15 & 2.00 \\
\hline
$x^{}_{K^0\pi^+\pi^-}$             & 0.23 & 2.23 \\
$y^{}_{K^0\pi^+\pi^-}$             & 2.49 & 4.73 \\
$|q/p|^{}_{K^0\pi^+\pi^-}$         & 0.00 & 4.73 \\
$\phi^{}_{K^0\pi^+\pi^-}$          & 0.67 & 5.39 \\
\hline
$R^{}_M(K^+\ell^-\nu)$           & 0.03 & 5.42 \\
\hline
$x^{}_{K^+\pi^-\pi^0}$             & 2.94 &  8.36 \\
$y^{}_{K^+\pi^-\pi^0}$             & 1.67 & 10.04 \\
\hline
$R^{}_M/y/R^{}_D/\sqrt{R^{}_D}\cos\delta$ (CLEOc) & 5.72 & 15.76 \\
\hline
$R^+/x'{}^{2+}/y'{}^+$ (\babar) & 2.74 & 18.50 \\
$R^-/x'{}^{2-}/y'{}^-$ (\babar) & 2.01 & 20.51 \\
$R^+/x'{}^{2+}/y'{}^+$ (Belle) & 3.72 & 24.23 \\
$R^-/x'{}^{2-}/y'{}^-$ (Belle) & 1.28 & 25.51 \\
$R^{}_D/x'{}^{2}/y'$     (CDF)   & 0.75 & 26.26 \\
\hline\hline
\end{tabular}
\end{center}
\end{table}

\begin{figure}
\includegraphics[width=80mm]{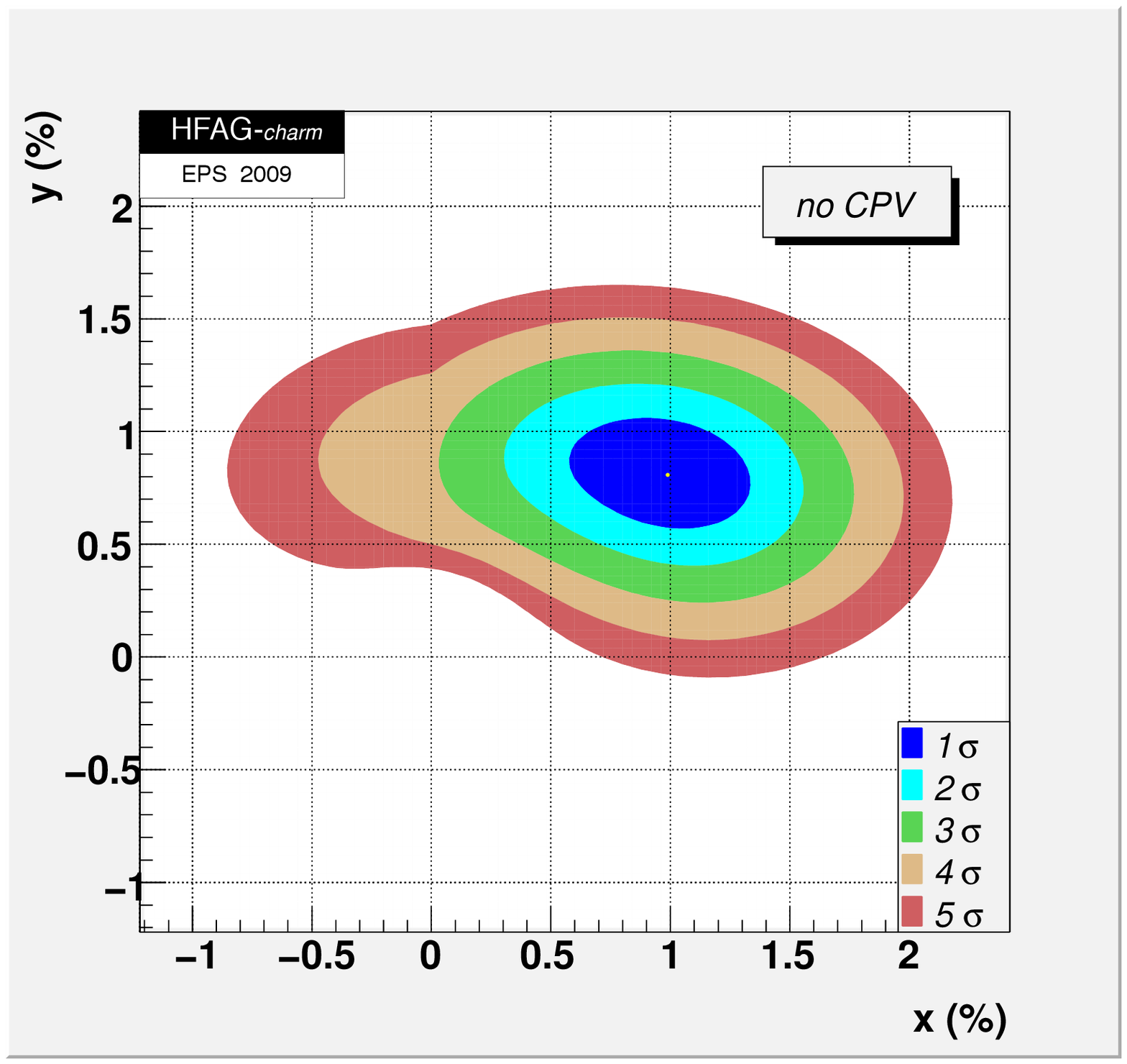}
\caption{Two-dimensional contours for mixing parameters $(x,y)$, for no \cpv. }
\label{fig:contours_ncpv}
\end{figure}

\begin{figure}
\vbox{
\includegraphics[width=80mm]{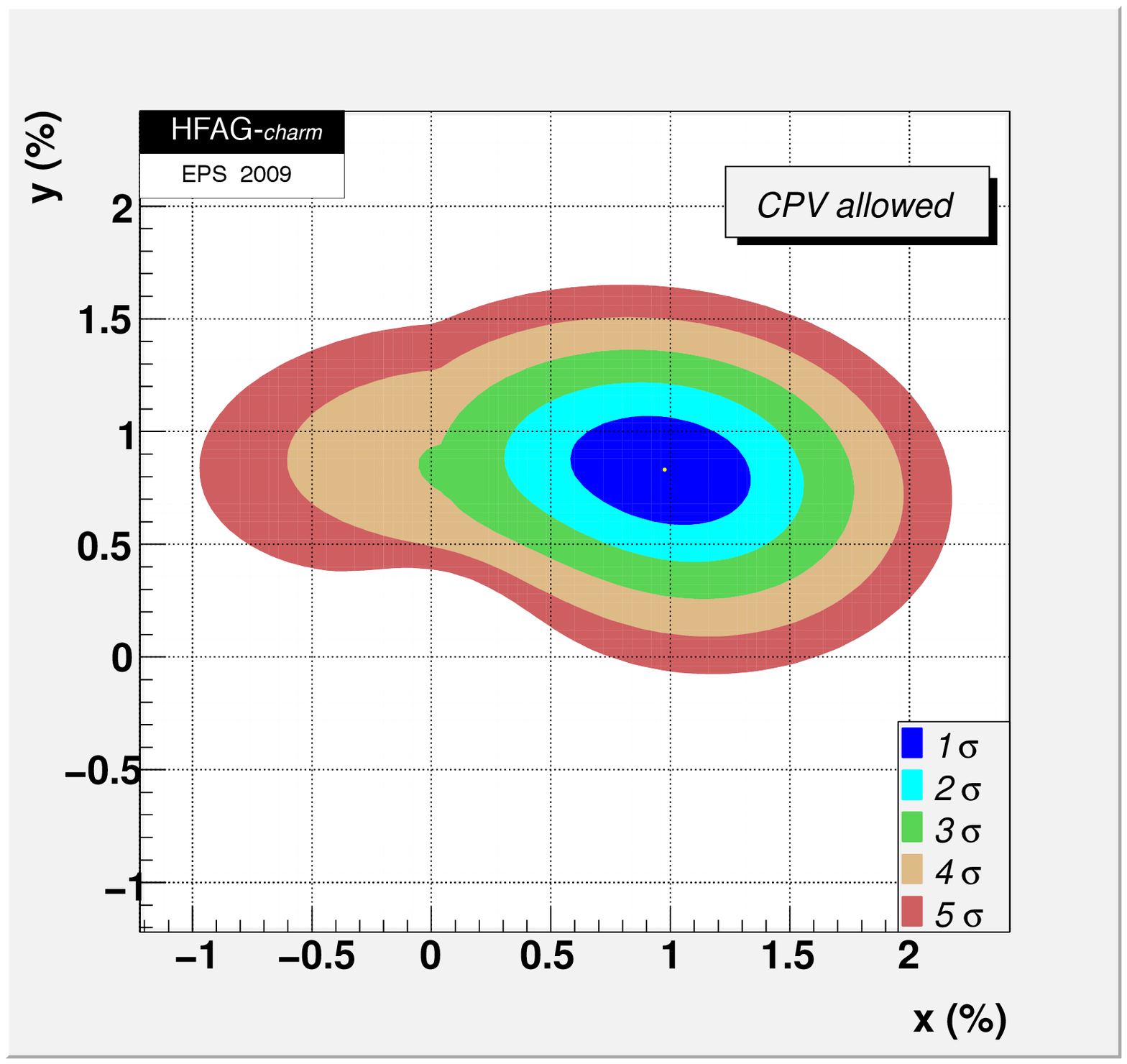}
\vskip0.15in
\includegraphics[width=80mm]{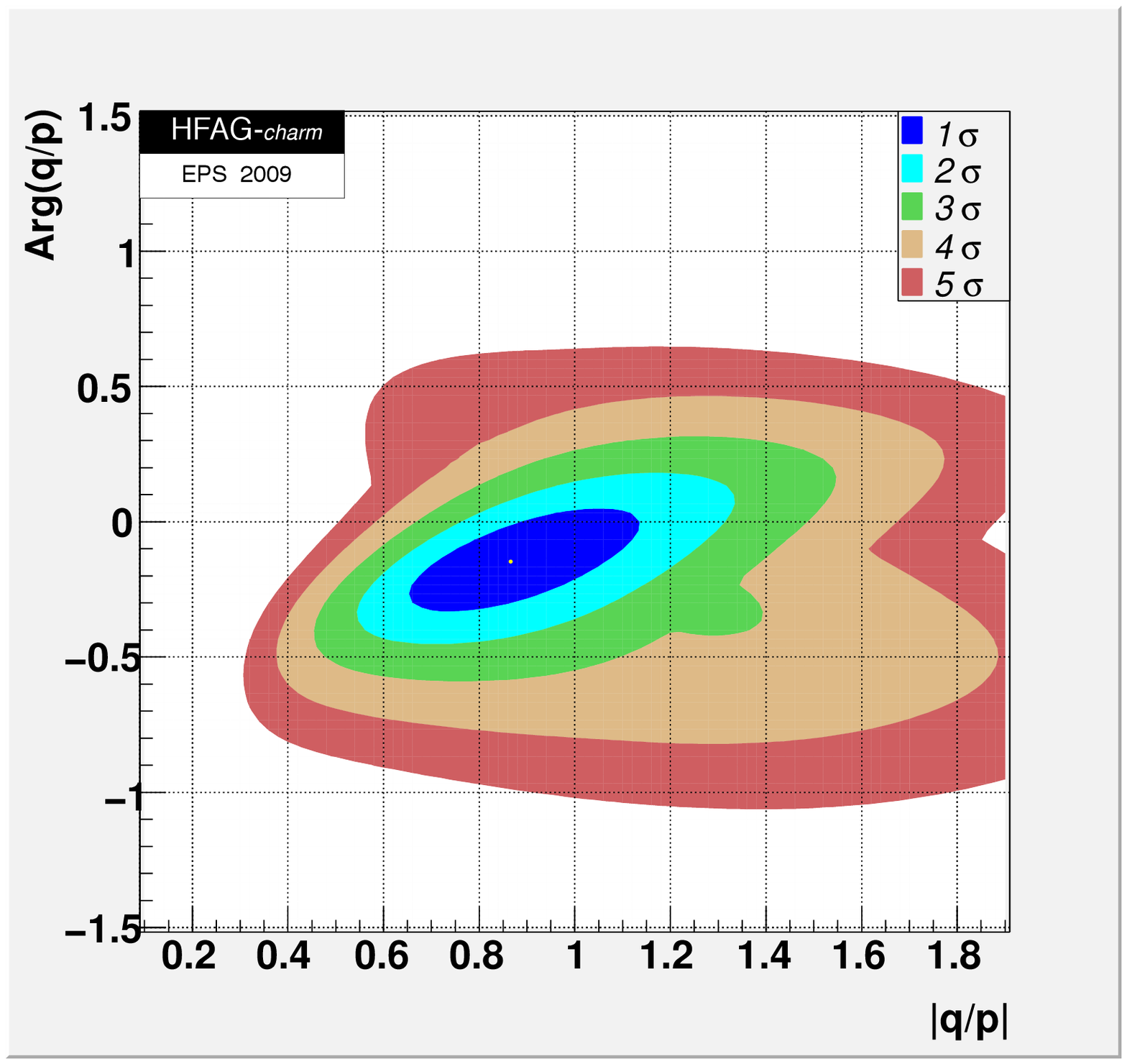}
}
\caption{\label{fig:contours_cpv}
Two-dimensional contours for parameters $(x,y)$ (top) 
and $(|q/p|,\phi)$ (bottom), allowing for \cpv.}
\end{figure}

Confidence contours in the two dimensions $(x,y)$ and
$(|q/p|,\phi)$ are obtained by letting, for any point in the
two-dimensional plane, all other fitted parameters take their 
preferred values. The resulting $1\sigma$-$5\sigma$ contours 
are shown in Fig.~\ref{fig:contours_ncpv} for the \cp-conserving
case, and in Fig.~\ref{fig:contours_cpv} for the \cpv-allowed 
case. The contours are determined from the increase of the
$\chi^2$ above the minimum value ($\chi^2_{\rm min}$). One 
observes that the $(x,y)$ contours for no-\cpv\ and for 
\cpv-allowed are almost identical. In the latter case, the $\chi^2$ 
at the no-mixing point $(x,y)\!=\!(0,0)$ is 110 units above the 
minimum value; this difference corresponds to a confidence level 
of $10.2\sigma$. Thus, no mixing is excluded at this high level.
In the $(|q/p|,\phi)$ plot, the no-\cpv\ point $(1,0)$ is within 
the $1\sigma$ contour; thus the data is consistent with 
\cp\ conservation.

\begin{figure*}
\hbox{\hskip0.50in
\includegraphics[width=72mm]{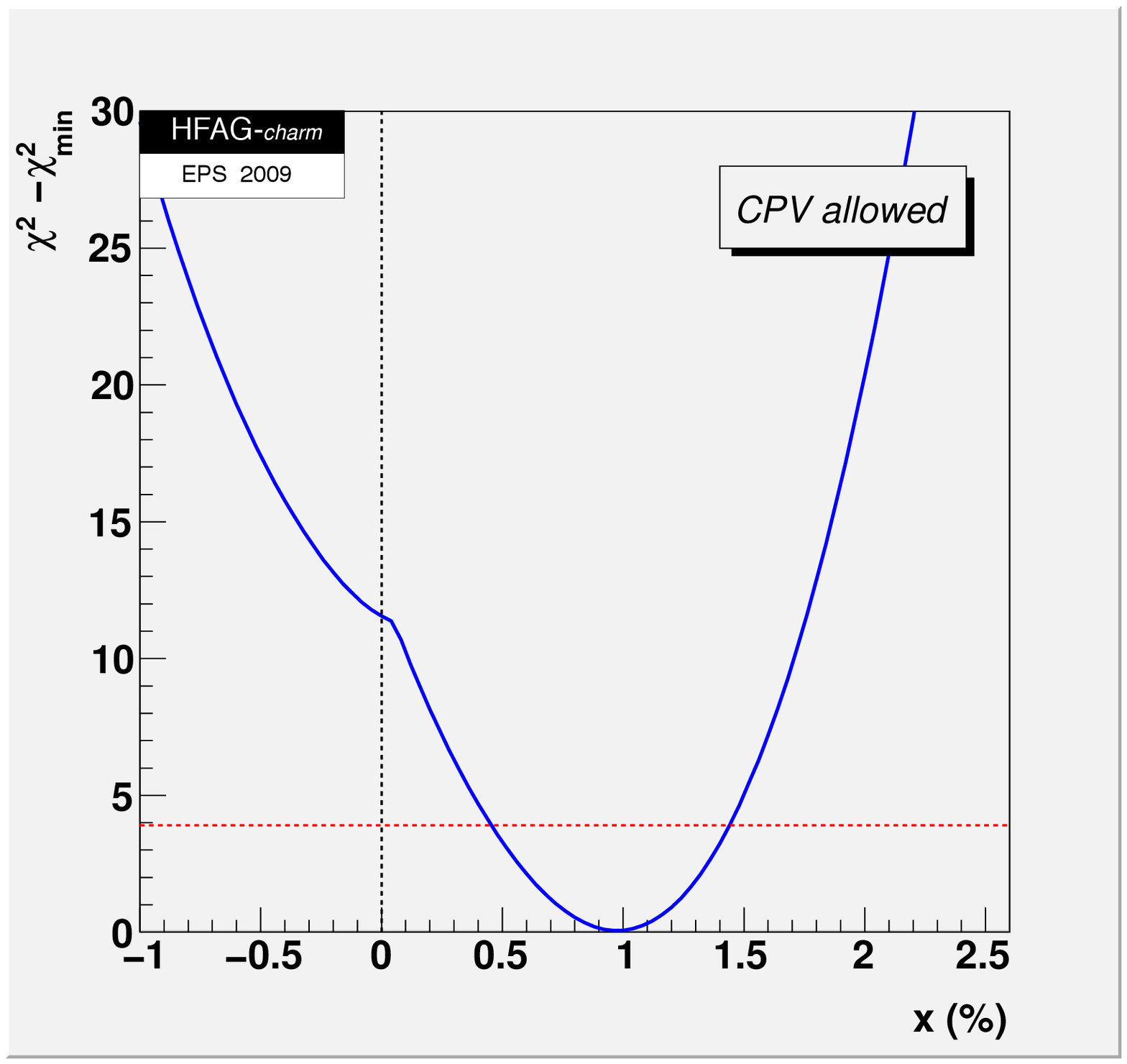}
\hskip0.20in
\includegraphics[width=72mm]{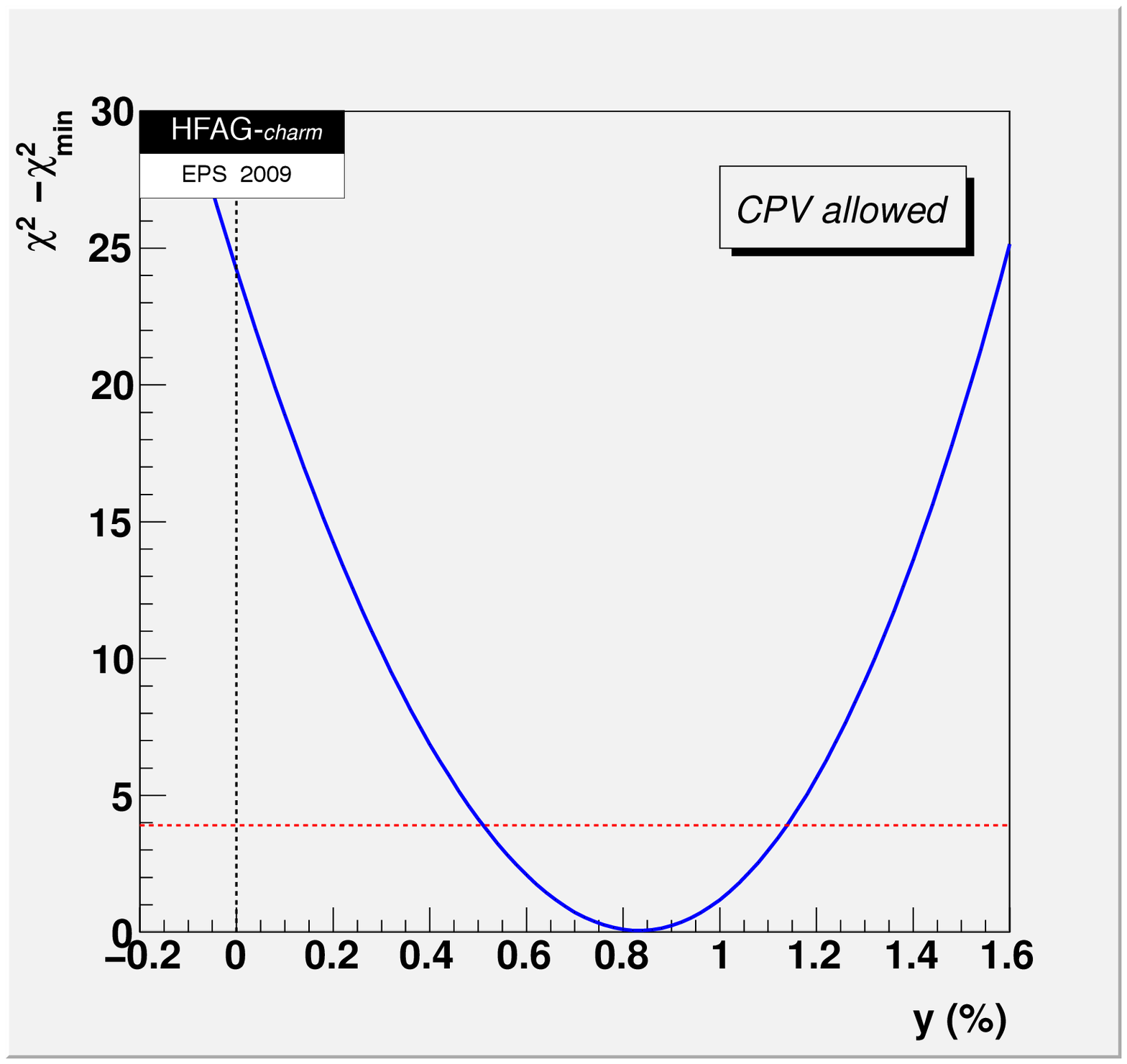}}
\hbox{\hskip0.50in
\includegraphics[width=72mm]{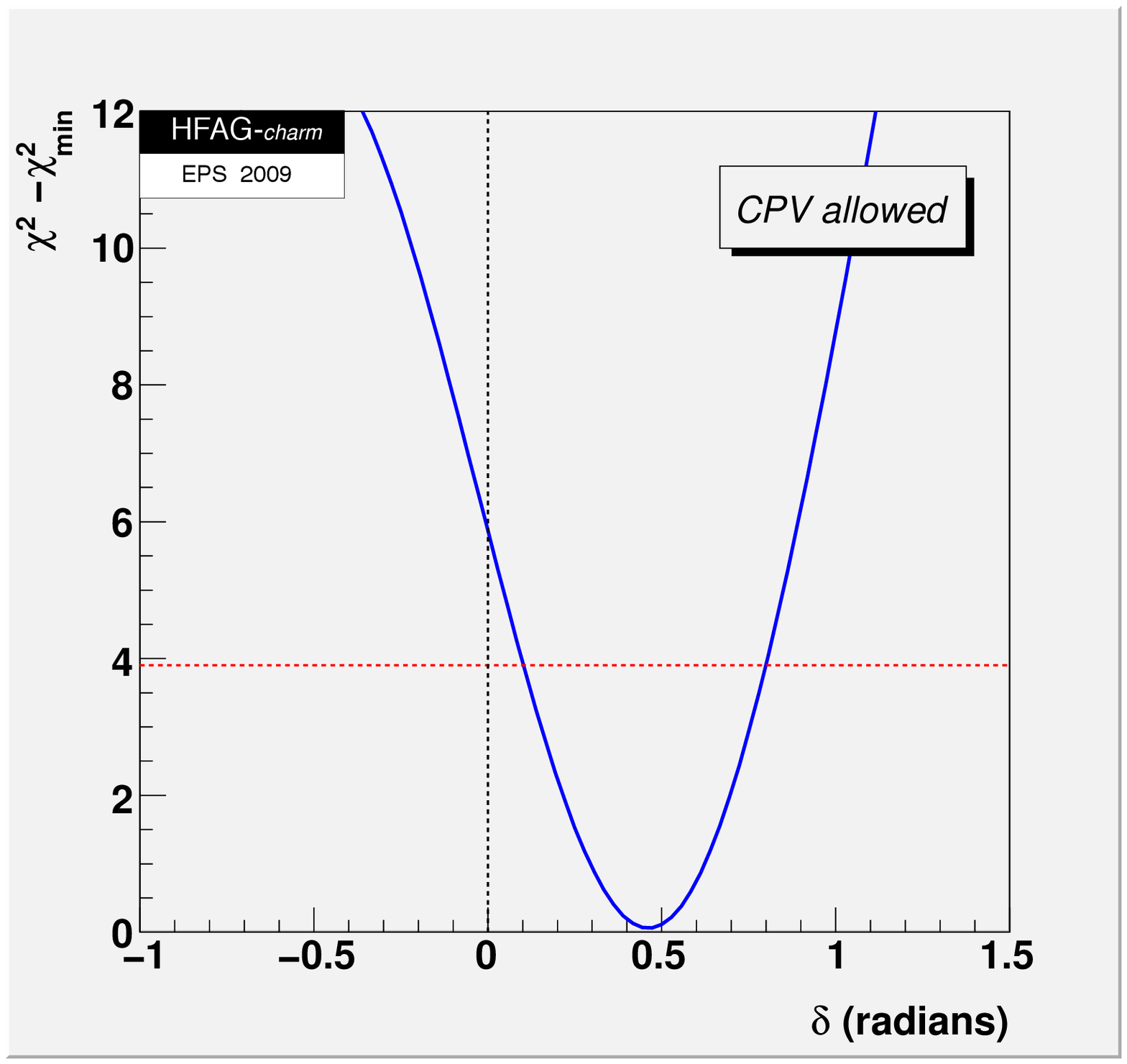}
\hskip0.20in
\includegraphics[width=72mm]{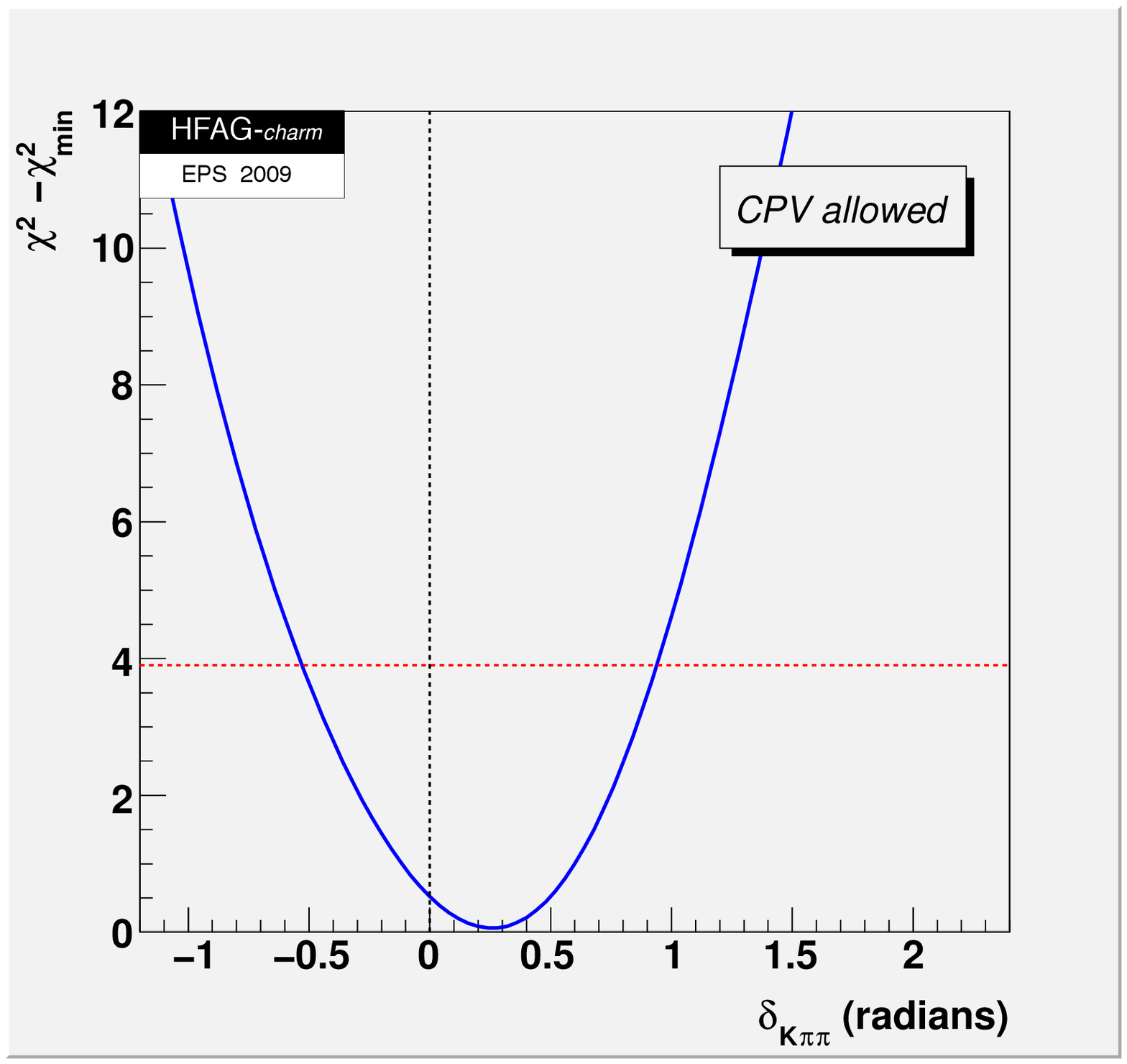}}
\hbox{\hskip0.50in
\includegraphics[width=72mm]{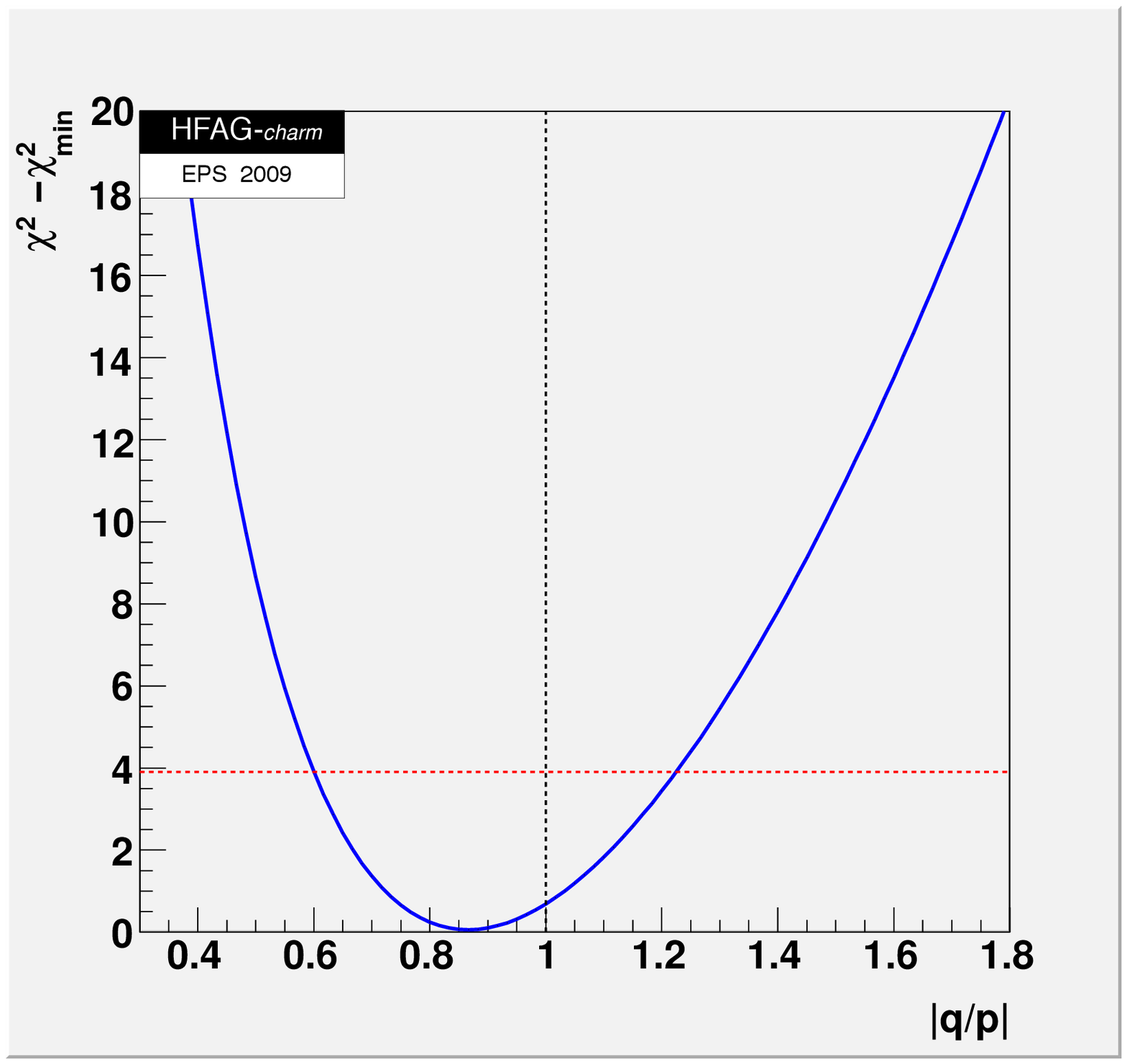}
\hskip0.20in
\includegraphics[width=72mm]{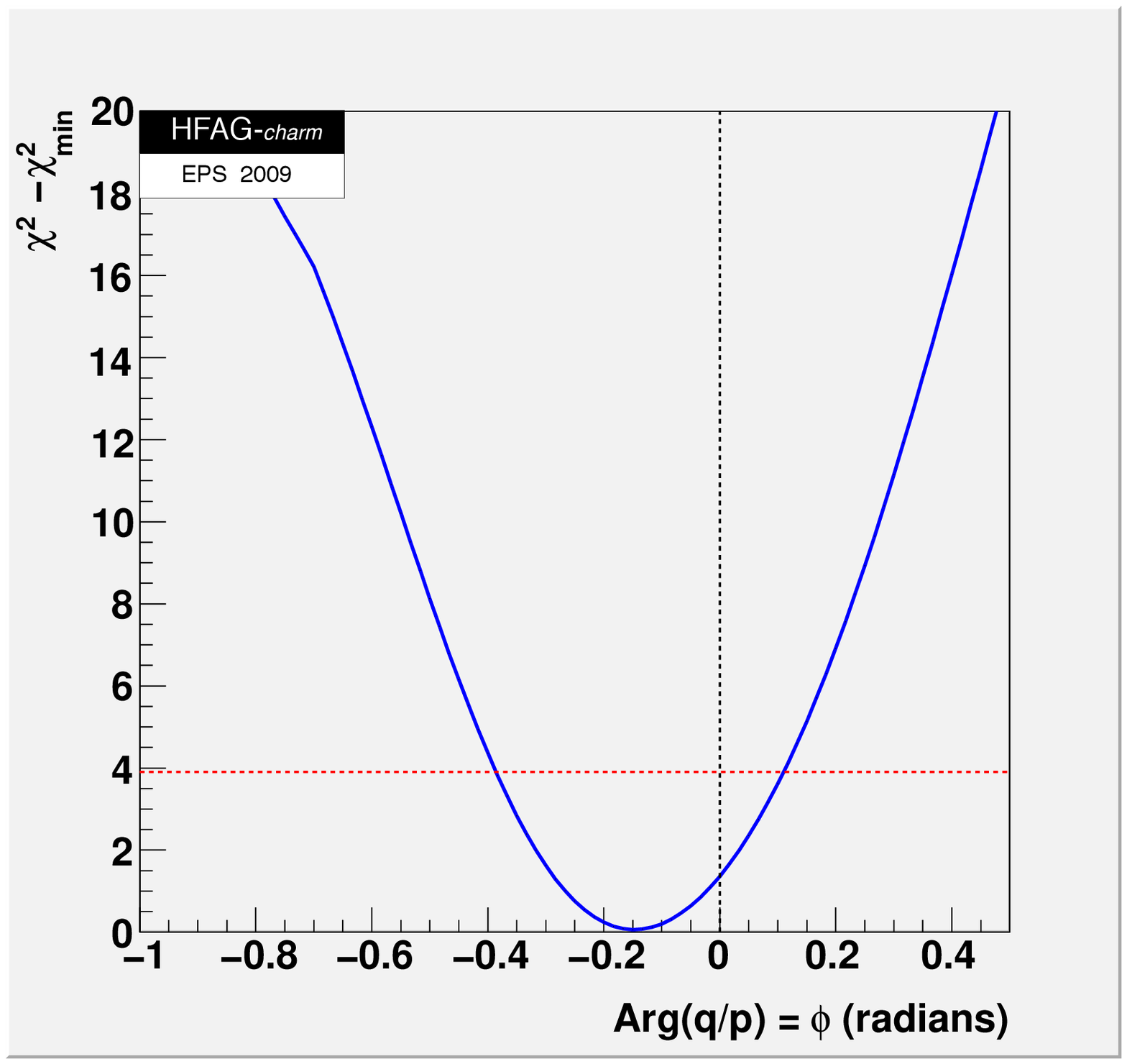}}
\vskip0.12in
\caption{\label{fig:1dlikelihood}
The function $\Delta\chi^2=\chi^2-\chi^2_{\rm min}$ for 
parameters $x,\,y,\,\delta,\,\delta^{}_{K\pi\pi},\,|q/p|$, 
and $\phi$. The points where $\Delta\chi^2=3.84$ (denoted 
by the dashed horizontal line) determine a 95\% C.L. interval.}
\end{figure*}

One-dimensional confidence curves for individual parameters 
are obtained by letting, for any value of the parameter, all 
other fitted parameters take their preferred values. The 
resulting functions $\Delta\chi^2=\chi^2-\chi^2_{\rm min}$ 
are shown in Fig.~\ref{fig:1dlikelihood}. The points where 
$\Delta\chi^2=3.84$ determine 95\% C.L. intervals for the 
parameters, as shown in the figure. These intervals are
listed in Table~\ref{tab:results}.

\section{\boldmath Summary}

We summarize the fit results listed in Table~\ref{tab:results} and shown 
in Figs.~\ref{fig:contours_cpv} and \ref{fig:1dlikelihood} as follows:
\begin{itemize}
\item the experimental data consistently indicate that $D^0$ 
mesons undergo mixing. The no-mixing point $(x,y)\!=\!(0,0)$ 
is excluded at $10.2\sigma$. The parameter $x$ differs from
zero by $3.2\sigma$, and the parameter $y$ differs from 
zero by $4.8\sigma$. The effect is presumably dominated 
by long-distance processes, which are difficult to calculate.
\item Since \ycp\ is positive, the \cp-even state is 
shorter-lived, as in the $K^0$-$\kbar$ system. However, 
since $x$ is also positive, the \cp-even state is heavier, 
unlike in the $K^0$-$\kbar$ system.
\item The strong phase difference $\delta$ is probably not small: 
the fitted value is $(26.4\,^{+9.6}_{-9.9})^\circ$.
\item There is no evidence yet for \cpv\ in the $D^0$-$\dbar$ 
system. Observing \cpv\ at the current level of sensitivity
would indicate new physics.
\end{itemize}

\section{Acknowledgments}

We thank the organizers of the CHARM 2009 workshop for 
excellent hospitality and for a stimulating scientific program.


\begin{thebibliography}{99}
\bibitem{hfag-charm} 
The group has since expanded its activities to include calculating 
WA values of hadronic branching fractions, semileptonic form factors, 
properties of excited $D^{**}_s$ states, and the $D^+_s$ decay 
constant~$f^{}_{D^{}_s}$.
\bibitem{charge-conjugates} 
Charge-conjugate modes are implicitly included unless noted otherwise.
\end{thebibliography}
\end{document}